\documentstyle[aps,prl]{revtex}

\begin{document}
\twocolumn[\hsize\textwidth\columnwidth\hsize
\csname @twocolumnfalse\endcsname
\title{$c$-Axis Twist Bi$_2$Sr$_2$CaCu$_2$O$_{8+\delta}$ 
Josephson Junctions: A New Phase-Sensitive Test of Order 
Parameter Symmetry}
\author{R. A. Klemm,$^1$ C. T. Rieck$^2$ and K. Scharnberg$^2$}
\address{$^1$Materials Science Division, Argonne National Laboratory,
Argonne, IL
60439 USA}
\address{$^2$Fachbereich Physik, Universit{\"a}t
Hamburg, Jungiusstra\ss e 11, D-20355 Hamburg, Germany}

\date{\today}
\maketitle

\begin{abstract}
   Li {\it et al.} found that the critical current 
density $J_c^J$ across atomically clean $c$-axis twist
 junctions of Bi$_2$Sr$_2$Ca Cu$_2$O$_{8+\delta}$ is the 
same as that of 
the constituent single crystal, $J_c^S$, independent of the twist 
angle 
$\phi_0$, even at $T_c$.   We investigated theoretically  if a 
$d_{x^2-y^2}$-wave order parameter might twist by mixing 
in $d_{xy}$ components, but find that such twisting cannot
 possibly explain the  data near to $T_c$.  
Hence,  the order parameter contains an $s$-wave component,
but  {\it not}  any 
$d_{x^2-y^2}$-wave component.  In addition, the $c$-axis 
Josephson tunneling is completely 
incoherent.  We also propose a $c$-axis junction tricrystal experiment which does not rely upon  expensive substrates.
\end{abstract}
\vskip0pt
\pacs{74.50.+r, 74.80.Dm, 74.72.Hs, 74.60.Jg}\vskip0pt
]
\narrowtext

\section{Experimental Introduction}
It has recently become possible to prepare extraordinarily 
perfect bicrystal Josephson junctions with Bi$_2$Sr$_2$CaCu$_2$O$_{8+\delta}$ (Bi2212). \cite{Li1,Li3} 
 These junctions are prepared by cleaving a very high quality 
single crystal of Bi2212 in the $ab$-plane, quickly examining 
the cleaved parts  under a microscope, rotating one part an 
arbitrary angle $\phi_0$ about the $c$-axis with respect to 
the other, placing them back together, and fusing them  by
heating  just below the melting point for 30 h. \cite{Li1,Li3} A schematic view of the resulting $c$-axis bicrystal is shown in 
Fig. 1.  By examining these bicrystals with high resolution transmission electron microscopy (HRTEM), electron energy-loss spectroscopy, and energy dispersive 
x-ray spectroscopy,   they were found to be atomically clean 
over the entire areas studied ($\approx10^2\mu{\rm m}^2$), and
 the periodic lattice distortion was atomically intact on each 
side of the twist junction. \cite{Li1}

 To each of the 12 bicrystals measured, six electrical leads 
were attached using Ag epoxy, two on opposite sides of 
each of the two constituent single crystals, and two across 
the bicrystal.  By 
applying the current $I$ across the central leads straddling
the bicrystal, and a voltage $V$ across two of the other
 leads, it was possible to measure the $I/V$ characteristics 
of 
the $c$-axis transport across each constituent single 
crystal, and 
across the twisted bicrystal junction in the same run.  
 The critical current $I_c$ was easily identified, as $V$
dropped by 5-8 orders of magnitude at a well-defined $I$ value, provided that the temperature $T$ was less than the transition temperature $T_c$. \cite{Li1,Li3}  For each bicrystal junction, 
they measured the critical current $I_c^J(T)$ and the junction 
area $A^J$.  Similarly, for each consituent single crystal, 
they measured the critical current $I_c^S(T)$ and the area 
$A^S$.  $I_c$ was often unmeasureable at low $T$, so 
comparisons 
of all 12 samples were made at $0.9T_c$. \cite{Li3}

Of the 12 bicrystal junctions, 7 were prepared with $40^{\circ}\le\phi_0\le50^{\circ}$, and one each at $0^{\circ}$ 
and $90^{\circ}$.  Although $I_c^S$ and $I_c^J$ at $T=0.9T_c$ 
varied from sample to sample, and the critical current densities 
$J_c^S=I_c^S/A^S$ and $J_c^J/A^J$ also varied from sample to 
sample, with only 
one exception (probably due to a sample that had 
weakly attached leads), {\it the ratio} $J_c^J/J_c^S$ {\it of critical current 
densities was the same} ($1.00\pm 0.06$) {\it for each sample!} 
 In a sample with $\phi_0=50^{\circ}$,  $J_c^J(T)/J_c^S(T)=1.0$ 
over the entire range 10 K$\le T <T_c$, and $I_c(T)/I_c(0)$ fit 
the  Ambegaokar-Baratoff curve. \cite{Li3,AB}  As discussed in 
the following, 
these results comprise very strong evidence for an $s$-wave component of the superconducting order parameter (OP) at and below $T_c$, and {\it cannot} be explained within a dominant
$d_{x^2-y^2}$-wave scenario.  

\section{Group Theory}

Although the crystal structure of Bi2212 is orthorhombic, the orthorhombic distortion is {\it different} from that of YBa$_2$Cu$_3$O$_{7-\delta}$, with different unit 
cell lengths $a$ and $b$ along the {\it diagonals} between the 
Cu-O bond directions in the CuO$_2$ planes.  In addition, there
 is an incommensurate lattice disortion ${\bf Q}=(0,0.212,1)$
 along one 
of these diagonals, the $b$-axis, which is clearly seen in 
the HRTEM pictures of the twist junctions. \cite{Li1}  Since 
${\bf Q}$ contains a $c$-axis component, only  the $bc$-plane is 
 a strict crystallographic mirror plane. \cite{Li1}  
Group theory
and Bloch's theorem dictate that the superconducting OP must 
reflect the crystal symmetry.  In Table 1, 
we have presented the allowable forms of the OP 
eigenfunctions 
 for Bi2212. \cite{KRS2} We presented 
both the angular momentum (fixed $k_F$, variable 
$\phi_{\bf k}$, with quantum numbers $\ell$)  and the 
nearly tetragonal lattice (variable $k_x, k_y$, with quantum numbers $n, m$) representations of  the OP eigenfunction 
forms. As indicated, 
the two OP eigenfunction forms are respectively 
even and odd with respect to reflections about the 
$bc$-plane 
(the $\sigma_b$ operation). \cite{Tinkham}  Thus, in 
Bi2212, $s$-wave and 
$d_{x^2-y^2}$-wave OP components are completely 
{\it incompatible}, and do not mix except possibly below 
a second
 (as yet unobserved) thermodynamic phase transition. 
\cite{KRS2}

\section{Lawrence-Doniach model}
Previously, we investigated whether it might be possible to 
explain the lack of any $\phi_0$-dependence of the $c$-axis 
critical current from a purely $d$-wave scenario. \cite{KRS}  We assumed that in the $n$th  layer, the dominant OP component 
was $d_{x^2-y^2}$, the amplitude $A_n$ of which became non-vanishing below $T_c=T_{cA}$.  By choosing the sub-dominant 
OP component to be $d_{xy}$ with amplitude $B_n$ and bare transition temperature $T_{cB}<T_{cA}$,  we considered
whether the overall OP could ``twist'' by mixing 
 $d_{x^2-y^2}$- and $d_{xy}$-wave components, to 
accommodate for the physical twist in the Josephson junction
between the adjacent layers $n=1$ and $n=-1$.  

\begin{table}
\vbox{\tabskip=0pt
\def\tablerule{\noalign{\hrule}}
\halign to245pt{\strut#&#\tabskip=0.39em plus1em&
\hfil#& &\hfil#&#&\hfil#&#&\hfil#&#&\hfil#&#&
\hfil#&#&\hfil#&#\hfil
\tabskip=0pt\cr
\noalign{\vskip5pt}\tablerule
\noalign{\vskip5pt}
 &&
\omit\hidewidth GT \hidewidth &&
\omit\hidewidth OP\hidewidth &&\omit\hidewidth $E$
 \hidewidth && \omit\hidewidth$\sigma_{d1}$\hidewidth
&&
\omit\hidewidth $\sigma_{d2}$\hidewidth
&&\omit\hidewidth $C_2$\hidewidth\cr
\noalign{\vskip5pt}
&&&&\hidewidth Eigenfunction \hidewidth&&&&
\omit\hidewidth ($\sigma_a$)\hidewidth &&\omit
\hidewidth ($\sigma_b$)\hidewidth &&
\cr
\noalign{\vskip5pt}\tablerule \noalign{\vskip5pt}
 &&$A_1$&&\hbox{$|s+d_{xy}\rangle$}
&&+1&&+1&&+1&&
+1&\cr\noalign{\vskip5pt}
&&&&\hbox{$ {\rightarrow\atop\ell}\>\>\tilde{a}_0+\sqrt{2}
\sum_{n=1}^{\infty}\{\tilde{a}_n\times$}&&&&&&&&&\cr
\noalign{\vskip5pt}
&&&&\hbox{$
\qquad\times\cos[2n(\phi_{\bf k}-\pi/4)]\}$}&&&&&&&&&\cr\noalign{\vskip5pt}
&&&&\hbox{$
{\rightarrow\atop{n,m}}\>\>\sum_{n,m=0}^{\infty}
\bigl\{[\tilde{a}_{nm}
+\tilde{a}_{mn}]\times$}&&&&&&&&&\cr\noalign{\vskip5pt}
&&&&\hbox{$\qquad\times\cos(nk_xa)\cos(mk_ya)$}
&&&&&&&&&\cr
\noalign{\vskip5pt}
&&&&\hbox{$
\qquad +[\tilde{c}_{nm}
+\tilde{c}_{mn}]\times$}&&&&&&&&&\cr\noalign{\vskip5pt}
&&&&\hbox{$\qquad\times\sin(nk_xa)\sin(mk_ya)\bigr\}$}
&&&&&&&&&\cr\noalign{\vskip5pt}
&&$A_2$&&\hbox{$|d_{x^2-y^2}+g_{xy(x^2-y^2)}
\rangle$}&&+1&&-1&&
-1&&+1&\cr\noalign{\vskip5pt}
&&&&\hbox{$ {\rightarrow\atop\ell}\>\> \sqrt{2}\sum_{n=1}^{\infty}\{\tilde{b}_n\times$}
&&&&&&&&&\cr
\noalign{\vskip5pt}
&&&&\hbox{$\qquad\times\sin[2n(\phi_{\bf
k}-\pi/4)]\}$}&&&&&&&&&\cr\noalign{\vskip5pt}
&&&&\hbox{$
{\rightarrow\atop{n,m}}\>\>\sum_{n,m=0}^{\infty}
\bigl\{[\tilde{a}_{nm}-\tilde{a}_{mn}]\times$}&&&&&&&&&\cr\noalign{\vskip5pt}
&&&&\hbox{$\qquad\times\cos(nk_xa)\cos(mk_ya)$}
&&&&&&&&&\cr
\noalign{\vskip5pt}
&&&&\hbox{$
\qquad +[\tilde{c}_{nm}-\tilde{c}_{mn}]\times$}&&&&&&&&&\cr\noalign{\vskip5pt}
&&&&\hbox{$\qquad\times\sin(nk_xa)\sin(mk_ya)\bigr\}$}
&&&&&&&&&\cr
 \noalign{\vskip5pt}\tablerule
\noalign{\vskip10pt}}}
\caption{Singlet superconducting OP eigenfunctions in the angular
momentum ($\ell$) and lattice ($n, m$) representations, their
 group
theoretic notations, and character table
for the orthorhombic point group $C_{2v}$ in the form
appropriate for
 BSCCO.   Although the
$\sigma_{d1}$
mirror plane symmetry is only approximate due to the
$c$-axis component of ${\bf Q}$, the $\sigma_{d2}$ mirror
plane
symmetry is robust in  most current samples.}\label{t5}
\end{table}

There are basically two distinct, relevant energy scales in this problem.  One is the relative amount of the two {\it incompatible}  $d$-wave components.  This is determined mainly by the different  
bare $T_c$ values, $T_{cA}$ and $T_{cB}$, arising from the pairing interactions.  Assuming there is only one observable zero-field superconducting 
phase transition at $T_c=T_{cA}$, then $T_{cB}<<T_{cA}$,  the correspondingly suppressed bulk $T_{cB}^{<}<T_{cB}$,  below which the OP is nodeless, and $|B_n|<<|A_n|$ for $n\rightarrow\pm\infty$.   This results in a strong 
locking of an anisotropic OP onto the lattice, with the anti-nodes of the purported $d_{x^2-y^2}$-wave OP component locking onto the Cu-O 
bond directions on each side of the twist junction.  For $\phi_0=45^{\circ}$, these OP eigenfunctions are thus orthogonal, 
and $I_c(45^{\circ})=0$.

The second energy scale is the strength of the Josephson coupling $\eta$ (and $\eta'$ across the twist junction) of the OP components between adjacent layers.  This gives rise to a finite $c$-axis coherence length, which diverges as $T\rightarrow T_c$,  
 allowing some {\it real} OP mixing (or twisting), suppressing $I_c(\phi_0)$ for $\phi_0\ne0$, as shown in Figs. 2 and 3.  
For strong interlayer coupling, $\eta\approx 1$, it is possible that $I_c(45^{\circ})\ne0$ for $T\approx 0.5T_c$, as pictured in Figs. 2 and 3.  However, Bi2212 is extremely anisotropic, and we thus 
expect $\eta\approx\eta'<<<1$.  In Fig. 3, we 
recalculated $I_c(\phi_0,T)$ for this case.  Clearly, for $\eta=\eta'=0.001$, $I_c(45^{\circ})\approx0$ for $T\ge 0.5T_c$.  Thus, it is {\it  extremely difficult, if not impossible} to explain the  data of 
Li {\it et al.} by assuming a dominant $d_{x^2-y^2}$-wave OP component with nodes at low $T$.\cite{Li3,KRS2,KRS}

\section{Conclusions}
We thus conclude that the superconducting OP in Bi2212 is the 
the group $A_1$, which contains the $s$-wave component, but 
does {\it not} contain any purported $d_{x^2-y^2}$-wave 
component near to $T_c$.  However, these experiments also 
provide information about the nature of the interlayer 
tunneling processes.   If  there 
were a substantial amount of coherent interlayer 
tunneling, then  the OP would be entirely isotropic $s$-wave 
in form.  For free-particle Fermi 
surfaces, this scenario is  possible, as rotated 
Fermi surfaces on opposite sides of the twist are degenerate.  However, Bi2212 is generally thought to have a tight-binding 
Fermi surface.  In this case, intertwist coherent tunneling is 
only possible for 
$\phi_0\approx 0, 90^{\circ}$, for which a finite fraction 
of the rotated Fermi 
surfaces are  degenerate.  Thus coherent tunneling 
would result in a larger $I_c(\phi_0)$ for $\phi_0=0^{\circ}, 90^{\circ}$ than for any 
other $\phi_0$ value,  contrary to experiment. \cite{Li3} Hence, 
we conclude that the interlayer tunneling is entirely incoherent, without any discernible interlayer forward scattering, even
between adjacent layers 
on the same side of the twist junction. \cite{KRS} 

Thus,  the OP eigenfunction is $|s+d_{xy}\rangle$ with GT notation $A_1$ (Table 1), which contains the $s$-wave component.  This OP eigenfunction could exhibit nodes, 
but it is {\it even} about reflections in the $bc$-mirror plane, and thus does not contain any amount of the odd $d_{x^2-y^2}$-wave 
OP component.  This conclusion is further supported by new 
$c$-axis Josephson junction experiments betweeen Bi2212 and
 Pb, which showed strong evidence for an $s$-wave component 
at low $T$. \cite{Kleiner}

\section{New tricrystal experiment proposal}

Since the $c$-axis junctions are qualitatively superior to the $ab$-plane thin film junctions, we propose a new tricrystal (or tetracrystal)
experiment  using $c$-axis junctions, as pictured
 in Fig. 4.  This experiment does not require any expensive substrates, and the grain boundaries are intrinsically far superior
 to those of the planar junctions. \cite{Li1,Kirtley} In addition, 
since $I_c$ for
the $c$-axis junctions is ordinarily much larger than for 
the $ab$-plane due to larger junction areas, it is much easier 
to satisfy the experimental requirement  $I_cL/\Phi_0>>1$, where $L$ is the induction of the ring and $\Phi_0$ is the flux quantum. 
\cite{KRS2}

\section{Acknowledgments}
Supported by  USDOE-BES  
through Contract No. W-31-109-ENG-38, by NATO through Collaborative Research Grant No. 960102, 
and by the DFG through the Graduiertenkolleg ``Physik nanostrukturierter Festk{\"o}rper.''  Send correspondence to {\tt klemm@anl.gov}.

\begin{figure}
\caption{Illustration of a $c$-axis twist junction with
 twist
angle $\phi_0$. }\label{fig1}
\end{figure}

\begin{figure}
\caption{Plot of $I_c(\phi_0)/I_c(0)$ for the case  of a dominant $d_{x^2-y^2}$-wave and subdominant
$d_{xy}$-wave OP, the relative amounts varying with layer 
index away
from the twist junction. [6] In these curves, 
$T_{cB}/T_{cA} = 0.2$, $T_{cB}^{<}/T_{cA} = 0.1304$,
$\epsilon/6\beta_A = 0.5$, $\delta/6\beta_A = 0.1$, and the 
$\eta=\eta'=1$ curves are solid.  At $t=T/T_{cA}=0.7$, curves for $\eta=1$ 
and various $\eta'$ values are shown. }\label{fig2}
\end{figure}

\begin{figure}
\caption{Plot of $I_c(\phi_0)/I_c(0)$ for the case considered in
[6] of a dominant $d_{x^2-y^2}$-wave and subdominant
$d_{xy}$-wave OP, the relative amounts varying with layer index away
from the twist junction.  
Curves for
$\eta=\eta'=0.1$ and $\eta=\eta'=0.001$ with $t= T/T_{cA} =0.99, 0.9, 0.5$ are presented. The other parameters are the same as in 
Fig. 2.}\label{fig3}
\end{figure}

\begin{figure}
\caption{ Proposed configuration of a $c$-axis
version
of the tricrystal ring experiment.  Dark crystal:  bottom.
Light crystal:
top.  Intermediate shading:  equal thickness crystals.
Arrows indicate the direction of a given single crystal axis.}\label{fig4}
\end{figure}

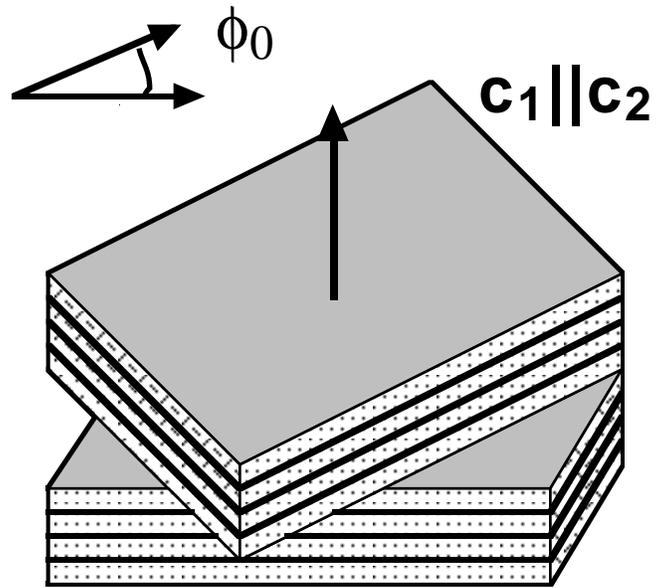
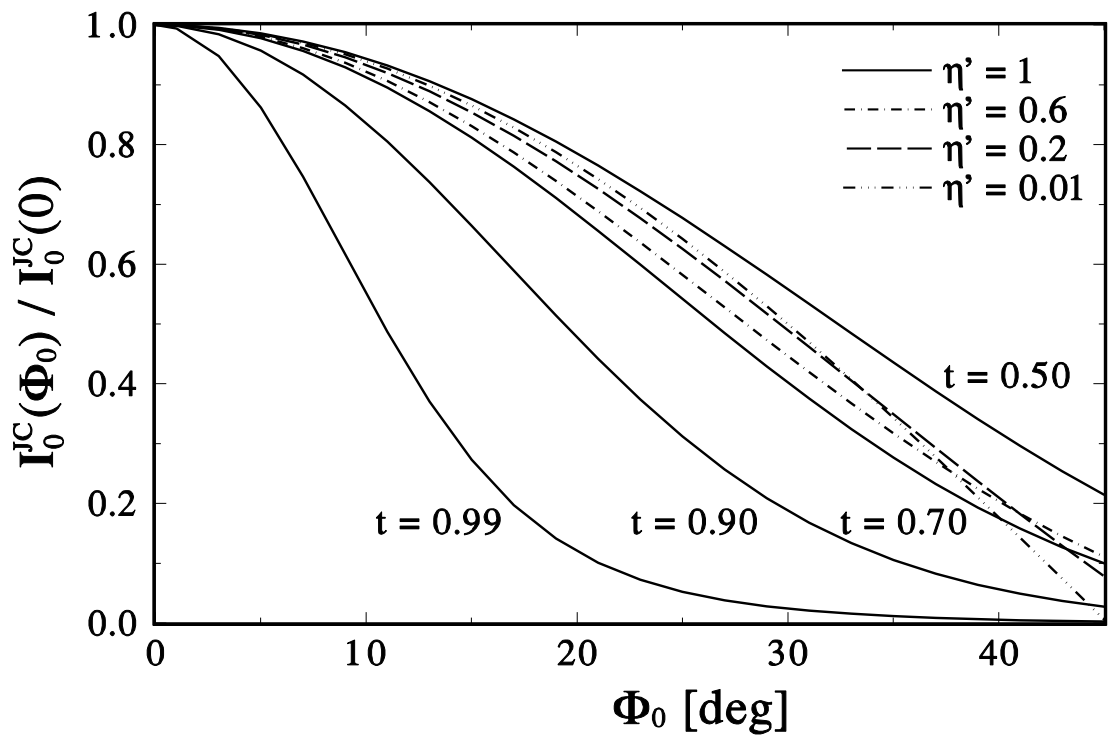
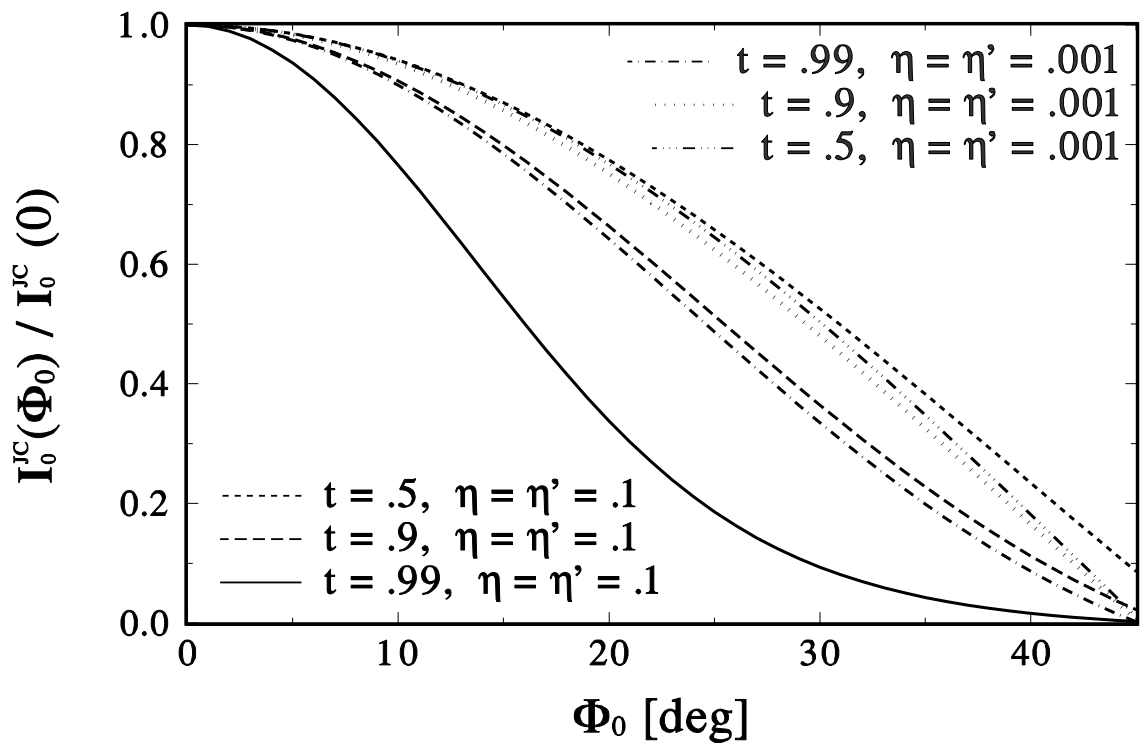
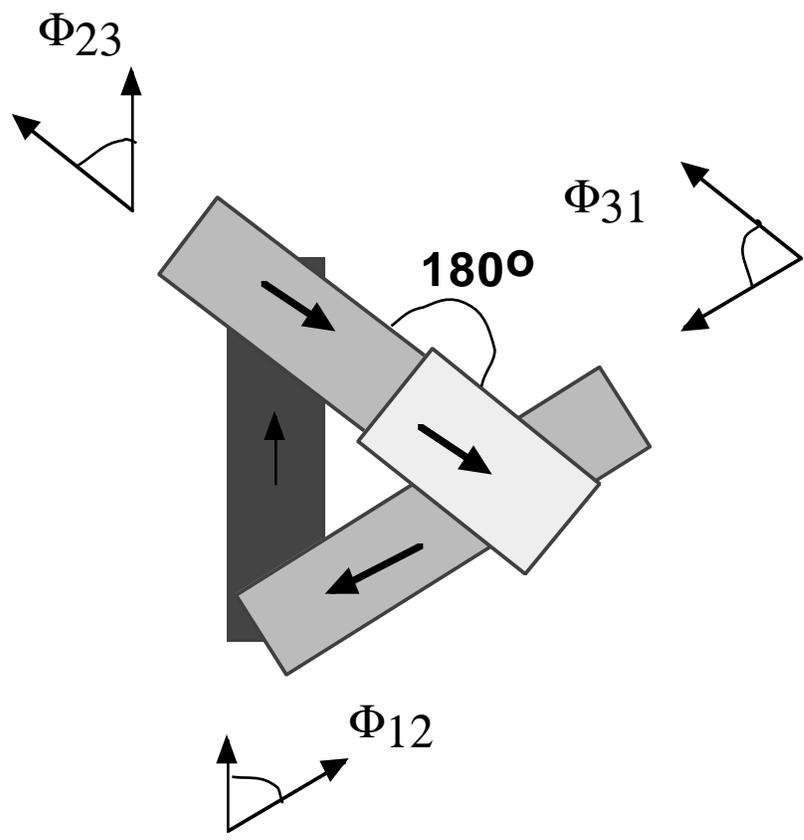
\end{document}